\documentclass[conference,a4paper]{IEEEtran}
\IEEEoverridecommandlockouts
\usepackage{cite}
\usepackage{amsmath,amssymb,amsfonts}
\usepackage{graphicx}
\usepackage{textcomp}
\usepackage{xcolor}
\usepackage{url}
\usepackage{booktabs}

\def\BibTeX{{\rm B\kern-.05em{\sc i\kern-.025em b}\kern-.08em
    T\kern-.1667em\lower.7ex\hbox{E}\kern-.125emX}}
\begin{document}

\title{Channel-Aware Selection of Folded Bloom Filters for Distributed Systems}

\author{
\IEEEauthorblockN{John Cartmell, Mihaela Cardei, Ionut Cardei}
\IEEEauthorblockA{
Department of Electrical Engineering and Computer Science\\
Florida Atlantic University\\
Boca Raton, FL, USA\\
jcartmell2023@fau.edu, mcardei@fau.edu, icardei@fau.edu\\
ORCID: 0000-0002-7014-4005, 0000-0003-2359-6196, 0009-0000-1050-768X
}
}

\maketitle

\begin{abstract}
Periodic Bloom-filter transmission can impose substantial overhead in communication-constrained distributed systems. Lossless compression preserves membership behavior but provides a single transmission size, whereas established OR folding produces smaller representations with higher false-positive rates (FPRs) while preserving the no-false-negative property.

This paper investigates channel-aware selection among OR-folded representations. The sender retains an unchanged canonical filter, constructs a catalog satisfying a maximum FPR, and selects the FPR-qualified representation with the largest retained length supported by the communication resources available at each reporting opportunity. Unlike folding driven principally by cardinality and false-positive constraints, selection is driven by time-varying communication conditions.

Using two phishing URL datasets, the framework is evaluated under Five-State Markov Capacity, Gilbert--Elliott burst-error, and Rayleigh block-fading models. Channel-aware folding improves communication efficiency and receiver freshness relative to complete-filter and lossless-compression baselines when communication opportunities vary substantially. Under the more favorable Gilbert--Elliott model, it remains competitive in efficiency while maintaining the freshest receiver state. These results show that FPR-qualified folded views provide useful transmission operating points when a recent lower-fidelity update is preferable to delaying a larger representation.
\end{abstract}


%%%%%%%%%%%%%%%%%%%%%%%%%%%%%%%%%%%%%%%%%%%%%%

\section{Introduction}
Bloom filters (BFs) are compact probabilistic structures for approximate set membership, providing efficient insertion and queries with no false negatives and a controllable false-positive rate (FPR)~\cite{bloom1970}. Their low memory and computational requirements support applications in networking, databases, security, and distributed systems~\cite{broder2004network}. As cloud, edge, and Internet-of-Things (IoT) systems increasingly exchange distributed state~\cite{shi2016edge}, BFs may require periodic transmission between components. This study extends our broader work on Bloom-filter encodings for machine-learning classification~\cite{cartmell2026bfml} and entropy-punctured regression~\cite{cartmell2026regression}.

When communication resources are constrained or time varying, complete-filter updates may be delayed or dropped, leaving stale receiver state. Lossless compressed BFs reduce transmission size without changing membership behavior~\cite{mitzenmacher2002compressed}, but produce a single payload determined by the filter contents. OR folding instead produces smaller, directly queryable views with increased FPR while preserving the no-false-negative property. Folding is established prior art~\cite{broder2004network,sailhan2012folding}; Sailhan and Stehr~\cite{sailhan2012folding} use highly factorable filters and integer-factor folding to adapt operative size principally to cardinality and prescribed FPR bounds.

\begin{figure*}[t]
\centering
\includegraphics[width=0.90\textwidth]{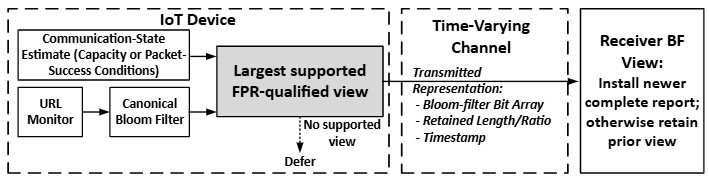}
\caption{Channel-aware folded-view selection, transmitted metadata, update deferral, and receiver-side version handling.}
\label{fig:system_overview}
\end{figure*}

This paper addresses the complementary problem of selecting a transmitted folded view according to an external, time-varying communication constraint. The sender retains the canonical filter, constructs a catalog of FPR-qualified folded views, and selects the view with the largest retained length supported at each reporting opportunity. Each update carries the view length and a timestamp; the receiver installs only newer, completely delivered reports and queries the selected view directly. The experiments hold the inserted set, canonical length, and hash count fixed to isolate channel-driven effects; this is an experimental control rather than a deployment requirement. Figure~\ref{fig:system_overview} illustrates the framework.

The research objective is to determine whether channel-aware selection among FPR-qualified folded views improves communication efficiency and receiver freshness relative to fixed complete-filter and lossless-compression baselines while preserving the no-false-negative property. Evaluation uses two phishing URL datasets and Five-State Markov Capacity, Gilbert--Elliott burst-error, and Rayleigh block-fading models.

The contributions are:
\begin{itemize}
\item an online policy for selecting an FPR-qualified OR-folded representation under time-varying communication constraints;
\item a fine-grained catalog of folded views derived from an unchanged canonical filter; and
\item an evaluation against complete-filter and lossless-compression baselines across two datasets and three channel models.
\end{itemize}

%%%%%%%%%%%%%%%%%%%%%%%%%%%%%%%%%%%%%%%%%%%%%%

\section{Related Work}
Mitzenmacher~\cite{mitzenmacher2002compressed} introduced compressed Bloom filters, which use entropy-based encoding to reduce transmission cost while permitting exact recovery of the original bit array. Bloom filters also serve as distributed summaries in web caching~\cite{fan2000counting} and set reconciliation~\cite{minsky2003set,eppstein2011strata}. These approaches do not select among multiple Bloom-filter views according to the conditions of an individual transmission opportunity.

OR-based folding is an established post-construction transformation~\cite{broder2004network,sailhan2012folding}. Broder and Mitzenmacher~\cite{broder2004network} describe halving a filter through bitwise OR and remapping queries to the reduced coordinate space, preserving the no-false-negative property while increasing FPR. Sailhan and Stehr~\cite{sailhan2012folding} select among integer-factor representations as cardinality changes to maintain a prescribed FPR range.

Other reduction methods use interleaving to select size from element count and maximum FPR~\cite{chen2020efficient}, truncation to allocate storage according to expected query distributions~\cite{mersy2024truncated}, or retained blocks to support flexible lengths~\cite{walther2026extending}. These methods respond principally to cardinality, FPR, query distribution, or storage constraints.

In contrast, the present work holds the inserted set, canonical filter, and hash count fixed while treating current communication conditions as the online selection variable. It selects the largest FPR-qualified fine-grained folded view, including non-integer-factor lengths, supported by each communication opportunity without resizing the canonical filter. To the best of our knowledge, time-varying communication conditions have not previously served as the online selection variable for Bloom-filter folding.

%%%%%%%%%%%%%%%%%%%%%%%%%%%%%%%%%%%%%%%%%%%%%%

\section{Methods}
\subsection{Bloom Filter Preliminaries}
\label{sec:bloom_filter_preliminaries}
A Bloom filter (BF) represents a set \(S\) of \(n\) elements using an \(m\)-bit array and \(k\) hash functions. Inserting an element sets the \(k\) corresponding bit positions to one, while a membership query returns positive when all corresponding positions are set. Classical Bloom filters permit false positives but not false negatives~\cite{bloom1970,broder2004network}. Their false-positive probability is approximated by
\begin{equation}
p_{\mathrm{FP}}\approx\left(1-e^{-kn/m}\right)^k,
\label{eq:bf_fpr}
\end{equation}
with the optimal number of hash functions given by
\begin{equation}
k=\frac{m}{n}\ln 2.
\label{eq:optimal_hashes}
\end{equation}
Equation~(\ref{eq:bf_fpr}) relates the FPR to \(m\), \(n\), and \(k\), while Equation~(\ref{eq:optimal_hashes}) gives the corresponding FPR-minimizing hash count.

\subsection{OR-Based Bloom Filter Folding}
OR-based folding is an established post-construction operation for reducing the length of a Bloom filter~\cite{broder2004network,sailhan2012folding}. Conventional one-half folding divides the bit array into two equal regions and combines corresponding positions using bitwise OR. Sailhan and Stehr further formalized integer-factor folding and unfolding for dynamically changing Bloom-filter cardinality and FPR requirements~\cite{sailhan2012folding}.

This work uses the same OR-folding principle but permits fine-grained retained lengths that need not be integer divisors of the canonical filter length. Let \(\mathbf{b}\in\{0,1\}^{m}\) denote the canonical Bloom filter, and let \(m'\leq m\) be the transmitted representation length. The folded representation \(\mathbf{b}'\in\{0,1\}^{m'}\) is constructed as
\begin{equation}
b'_i=\bigvee_{\substack{0\leq j<m\\j\bmod m'=i}} b_j,
\qquad 0\leq i<m'.
\label{eq:folding}
\end{equation}
Thus, Equation~(\ref{eq:folding}) maps every canonical position deterministically into the folded coordinate space and combines positions sharing the same remainder using bitwise OR. When \(m'=m/2\), it reduces to one-half folding. Values such as \(m'=0.95m\), \(0.90m\), or \(0.75m\) provide finer-grained communication operating points.

\begin{figure}[t]
\centering
\includegraphics[width=0.90\columnwidth]{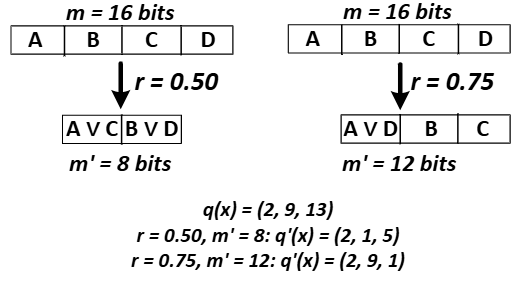}
\caption{OR-based folding of a 16-bit canonical Bloom filter at \(r=0.50\) and \(r=0.75\). The example shows how the canonical query positions for one element are mapped into each folded representation.}
\label{fig:folding}
\end{figure}

Figure~\ref{fig:folding} illustrates the operation for two retained lengths. The canonical filter remains available at the sender, allowing multiple folded representations to be generated or cached without reinserting the original set elements.

\subsection{Membership Queries After Folding}
For non-integer folding ratios, membership queries must preserve the same two-stage coordinate mapping used to construct the folded representation. Let \(H_i(x)\) denote the output of the \(i\)-th hash function. Its canonical position is
\begin{equation}
q_i(x)=H_i(x)\bmod m,
\label{eq:canonical_query}
\end{equation}
and its corresponding folded position is
\begin{equation}
q'_i(x)=q_i(x)\bmod m'.
\label{eq:folded_query_mapping}
\end{equation}
The folded membership query is therefore
\begin{equation}
\mathrm{BF}'(x)=
\bigwedge_{i=1}^{k} b'_{q'_i(x)}.
\label{eq:folded_membership}
\end{equation}

Equations~(\ref{eq:canonical_query})--(\ref{eq:folded_membership}) define the complete folded-query mapping. Using zero-based positions, the example in Figure~\ref{fig:folding} has \(q(x)=(2,9,13)\), which maps to \(q'(x)=(2,1,5)\) when \(m'=8\) and to \(q'(x)=(2,9,1)\) when \(m'=12\).

The two-stage mapping is important when \(m'\) is not a divisor of \(m\), because \(\big(H_i(x)\bmod m\big)\bmod m'\) is not generally equal to \(H_i(x)\bmod m'\). For every inserted element, the bit at each canonical position \(q_i(x)\) contributes by OR to the corresponding folded position \(q'_i(x)\). Folding therefore preserves the no-false-negative property while potentially increasing false positives, consistent with established OR-folding behavior~\cite{broder2004network,sailhan2012folding}.

\subsection{Folding Ratio and False Positive Behavior}
Let
\begin{equation}
r=\frac{m'}{m}
\label{eq:folding_ratio}
\end{equation}
denote the retained-length ratio. Lower values of \(r\) merge more canonical positions and increase folded-filter occupancy. If \(p_b\) is the probability that a canonical bit is set and approximately \(f=m/m'\) canonical positions contribute to each folded position, the resulting occupancy can be approximated by
\begin{equation}
p'_b\approx 1-(1-p_b)^f,
\label{eq:folded_occupancy}
\end{equation}
giving the approximate false-positive probability
\begin{equation}
p'_{\mathrm{FP}}\approx(p'_b)^k.
\label{eq:folded_fpr}
\end{equation}
Equations~(\ref{eq:folding_ratio})--(\ref{eq:folded_fpr}) provide intuition rather than an exact prediction because modulo folding can produce unequal preimage sizes when \(m'\) does not divide \(m\), and the merged positions are not strictly independent. The experiments therefore measure the FPR of every candidate representation directly.

\subsection{Channel-Adaptive Transmission}
The channel-adaptive framework is illustrated in Figure~\ref{fig:system_overview}. Each sender retains a complete canonical Bloom filter and generates or caches a catalog of representations at the retained-length ratios supported by the system. The catalog includes the complete canonical representation and progressively smaller OR-folded representations. Each candidate is associated with its retained length and measured FPR, and representations exceeding the application-defined maximum FPR are excluded.

At each reporting opportunity, the sender obtains an estimate of the current communication conditions from the underlying communication system and selects the highest-fidelity FPR-qualified representation supported under those conditions. For a capacity-limited channel, this is the representation with the largest retained length whose total transmitted size, including the Bloom-filter payload and packet headers, fits within the available capacity. Favorable conditions may support the complete canonical representation or a lightly folded view, whereas constrained conditions may require a smaller representation. If no qualified representation can be delivered, the update is deferred.

For a loss-based channel without an explicit byte-capacity limit, selection instead considers the probability of complete-report delivery. Smaller representations require fewer packets and may therefore have a greater delivery probability. The specific capacity- and loss-based policies used in the experiments are described in Section~IV.

Each transmitted update contains the selected Bloom-filter representation, its retained length \(m'\), and a timestamp that serves as its report version. The receiver installs only a completely delivered update newer than its currently stored view and queries it using the coordinate mapping in Equation~(\ref{eq:folded_query_mapping}). Stale, duplicate, or incomplete reports are not installed; after a failed or deferred update, the receiver retains its previous valid view. Reconstruction of the canonical filter is neither required nor attempted, and the canonical filter retained by the sender is not modified by selection or transmission. The experiments model complete-report delivery within individual reporting opportunities and do not separately model packet reordering.

To isolate the effect of time-varying communication conditions, the experiments hold the inserted set, canonical filter length, hash count, and candidate catalog constant. Each reporting opportunity is treated as a new timestamped update, while only the channel conditions and selected transmitted view vary. This controlled design ensures that differences in delivery, communication efficiency, and receiver age arise from the channel and representation-selection policy rather than from changes in Bloom-filter contents or cardinality.

%%%%%%%%%%%%%%%%%%%%%%%%%%%%%%%%%%%%%%%%%%%%%%

\section{Experimental Setup}
\label{sec:experimental_setup}
The evaluation uses two phishing URL datasets, twelve retained-length ratios, four lossless-compression baselines, and three time-varying channel models. Table~\ref{tab:parameters} summarizes the principal parameters. Within each dataset, model, and repetition, all strategies use the same channel trace.

\subsection{Datasets}
The PhiUSIIL~\cite{prasad2024phiusiil} and URL-Phish~\cite{minh2026urlphish} datasets are evaluated using 5,000 malicious URLs for insertion and 50,000 disjoint benign URLs as absent queries. Every positive response to a benign query is therefore a false positive. This models a distributed security application in which a sender reports known malicious URLs and a receiver queries unseen URLs.

The inserted set remains fixed within each run, holding representation size and error behavior constant to isolate the effects of channel conditions and transmission policy. This experimental control does not require static contents in deployment.

\subsection{Bloom Filter Construction}
For each dataset, the classical filter length \(m\) is selected for \(n=5{,}000\) elements and a target FPR of 1\%, then rounded upward to a complete byte. The classical baseline uses \(k_{\mathrm{classical}}=\operatorname{round}((m/n)\ln 2)\), giving an expected occupancy near 50\%.

The sparse canonical filter uses the same \(m\), with its hash count selected for a target occupancy \(\rho=0.25\):
\begin{equation}
k_{\mathrm{sparse}}=
\operatorname{round}\left(
-\frac{m}{n}\ln(1-\rho)
\right).
\label{eq:sparse_hash_count}
\end{equation}
Equation~(\ref{eq:sparse_hash_count}) produces three hash functions. One-half folding then increases the approximate occupancy from 0.25 to 0.4375 and gives an approximate FPR of 0.084. Thus, the 25\% target provides folding headroom while retaining sparsity for lossless compression; it is a controlled operating point rather than a universal optimum. The classical and sparse complete filters are separate baselines, and all folded and compressed representations derive from the sparse filter.

Hash positions use SHA-256 double hashing. Two domain-separated digests provide \(h_1(x)\) and \(h_2(x)\), with canonical positions
\begin{equation}
g_i(x)=\left(h_1(x)+i h_2(x)\right)\bmod m,
\qquad i=0,\ldots,k-1,
\label{eq:double_hashing}
\end{equation}
following Kirsch and Mitzenmacher~\cite{kirsch2008less}. After Equation~(\ref{eq:double_hashing}) generates the canonical positions, Equation~(\ref{eq:folding}) is applied to the sparse filter. Candidate lengths are rounded to complete bytes and evaluated using the inserted and absent-query sets.

Only representations with no observed false negatives and an empirical FPR no greater than 10\% enter the adaptive catalog. This illustrative ceiling admits moderate folding while excluding representations approaching saturation; a deployment would set it according to downstream false-positive cost.

\begin{table}[t]
\centering
\caption{Principal experimental parameters.}
\label{tab:parameters}
\begin{tabular}{@{}ll@{}}
\toprule
Parameter & Value \\
\midrule
Random seed & 42 \\
Inserted/absent URLs & 5,000/50,000 \\
Classical target FPR & 1\% \\
Classical/sparse occupancy & 50\%/25\% \\
Adaptive maximum FPR & 10\% \\
Retained-length ratios &
$\{1.00,0.95,0.90,0.85,0.80,0.75,$ \\
& $0.70,0.60,0.50,0.40,0.30,0.25\}$ \\
Transmission-opportunity duration & 10 ms \\
Packet payload/header & 480/32 bytes \\
Transmission opportunities per trace & 500 \\
Repetitions & 20 \\
\bottomrule
\end{tabular}
\end{table}

\subsection{Compression and Transmission Baselines}
The sparse canonical bit array is compared with gzip~\cite{deutsch1996deflate}, zlib~\cite{deutsch1996zlib}, bzip2~\cite{seward1998bzip2}, and LZMA~\cite{pavlov2024lzma}. Each uses compression level or preset 9, the highest numbered standard setting supported by the implementation. This provides a conservative size comparison, although the meaning and computational cost of level 9 are algorithm specific. Because these methods are lossless, they retain the sparse filter's measured FPR and false-negative rate. Uncompressed classical and sparse filters are also included.

Payloads are segmented into 480-byte packets with 32-byte headers. For \(s\) payload bytes, the total transmitted wire size is
\begin{equation}
W(s)=s+32\left\lceil\frac{s}{480}\right\rceil.
\label{eq:wire_size}
\end{equation}
For the Five-State Markov and Rayleigh models, each interval is a 10-ms transmission opportunity with a total byte budget. The 10-ms value specifies the channel-sampling resolution and Rayleigh block duration, not an application reporting rate. Each opportunity produces a newly timestamped report; delivery requires the wire size in Equation~(\ref{eq:wire_size}) to fit the available budget.

\subsection{Adaptive Transmission Policy}
For the Five-State Markov and Rayleigh models, the sender selects the valid representation with the largest retained-length ratio whose complete wire size fits the current capacity. If none fits, the update is deferred. A fixed baseline succeeds only when its complete representation fits.

The Gilbert--Elliott model represents packet loss rather than byte capacity. The sender selects the largest valid representation whose expected complete-report success probability is at least 0.50. If none qualifies, it selects the representation with the highest success probability. Delivery requires every packet to succeed. In every model, the canonical filter, inserted set, and candidate catalog remain fixed; only the transmitted representation changes.

\subsection{Channel Models}
The Five-State Markov model uses capacity ratios \(\{0.45,0.65,0.80,0.92,1.05\}\) relative to the complete sparse-filter wire size and begins in the middle state. Interior states remain unchanged with probability 0.60 and move to either adjacent state with probability 0.20 each; boundary probabilities are adjusted to remain within the five-state range.

The Gilbert--Elliott model begins in the Good state, with transition probabilities \(P(G\rightarrow B)=0.08\) and \(P(B\rightarrow G)=0.25\)~\cite{gilbert1960,elliott1963}. Packet-success probabilities are 0.995 and 0.90 in the Good and Bad states, respectively. The state remains constant within a report and transitions between opportunities; packet outcomes are conditionally independent, and delivery requires every packet to succeed.

For Rayleigh block fading, an independent unit-mean exponential power gain is generated for every opportunity. The average SNR is 10 dB, and capacity is
\begin{equation}
C_t=BT\log_2(1+\gamma_t),
\label{eq:rayleigh_capacity}
\end{equation}
where \(T=10\) ms and \(\gamma_t\) is the instantaneous SNR~\cite{shannon1948}. Capacity is expressed in byte-equivalent units, with \(B\) normalized so that Equation~(\ref{eq:rayleigh_capacity}) at the average SNR equals the complete sparse-filter wire size. The three models represent correlated capacity variation, burst-dependent packet loss, and independent multipath fading rather than a particular radio standard.

\subsection{Evaluation Metrics}
Bloom-filter performance is measured by empirical FPR versus retained-length ratio. Every candidate is verified to have no observed false negatives, and only candidates with FPR no greater than 10\% are eligible for adaptive selection.

Communication efficiency is measured as successful report deliveries per one million attempted wire bytes. Receiver age is the number of transmission-opportunity intervals since the most recent successful delivery. Figure~\ref{fig:folding_tradeoff} reports one canonical-filter realization per dataset using seed 42, without across-seed averaging or error bars. Figures~\ref{fig:communication_efficiency} and~\ref{fig:receiver_age} report means and 95\% confidence intervals over 20 paired traces of 500 opportunities, with all strategies using the same trace within each repetition.

%%%%%%%%%%%%%%%%%%%%%%%%%%%%%%%%%%%%%%%%%%%%%%

\section{Experimental Results}
\label{sec:results}
The evaluation addresses three questions: (1) how folding affects representation size and false positive rate, (2) whether channel-aware selection improves communication efficiency, and (3) whether any improvement results in fresher receiver state.

\subsection{Bloom Filter Folding Tradeoff}
\begin{figure}[t]
    \centering
    \includegraphics[width=0.90\columnwidth]{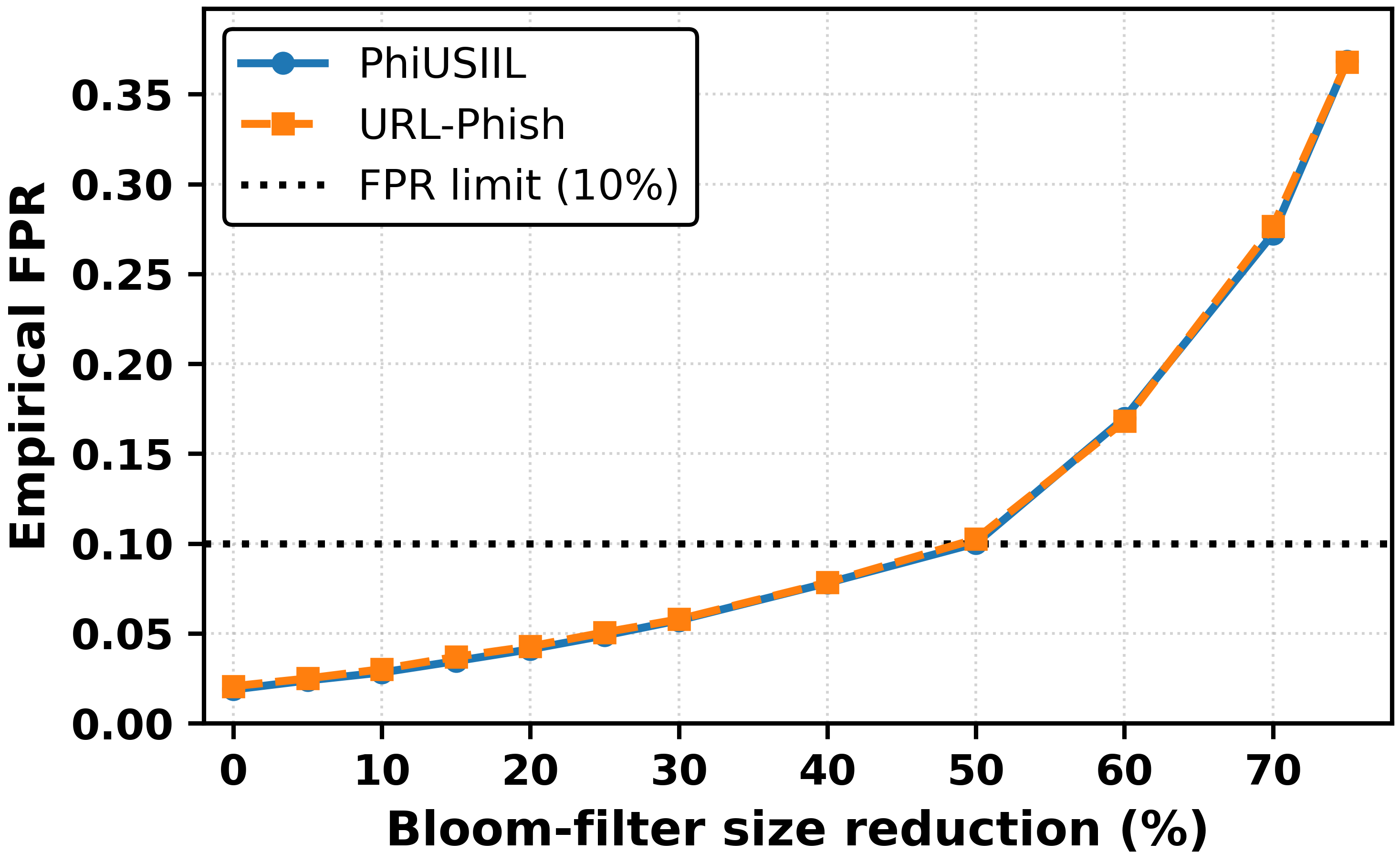}
    \caption{Empirical false-positive rate versus Bloom-filter size reduction, \(100(1-r)\), where \(r=m'/m\). Each curve represents one canonical-filter realization for its dataset using the fixed random seed of 42; no averaging across filter-construction seeds was performed.}
    \label{fig:folding_tradeoff}
\end{figure}

Figure~\ref{fig:folding_tradeoff} shows the empirical false positive rate as the retained-length ratio \(r=m'/m\) decreases. Figure~\ref{fig:folding_tradeoff} characterizes the exact canonical-filter realizations used in the subsequent channel experiments; because each dataset uses one fixed-seed realization, no across-seed error bars are shown.

The two datasets produce nearly identical false positive rates across the evaluated ratios. Because both filters contain the same number of inserted elements and use the same construction parameters, this similarity indicates that folding behavior is governed primarily by filter occupancy and the deterministic position mapping rather than by one particular URL collection.

The false positive rate increases gradually under moderate folding and then rises more rapidly as the reduced representation approaches saturation. Approximately 50\% size reduction remains near the 10\% maximum FPR used by the adaptive policy. The resulting valid representations provide multiple communication-cost and accuracy operating points; representations exceeding the FPR limit are excluded from adaptive selection.

\subsection{Communication Efficiency}
\begin{figure}[t]
    \centering
    \includegraphics[width=0.90\columnwidth]{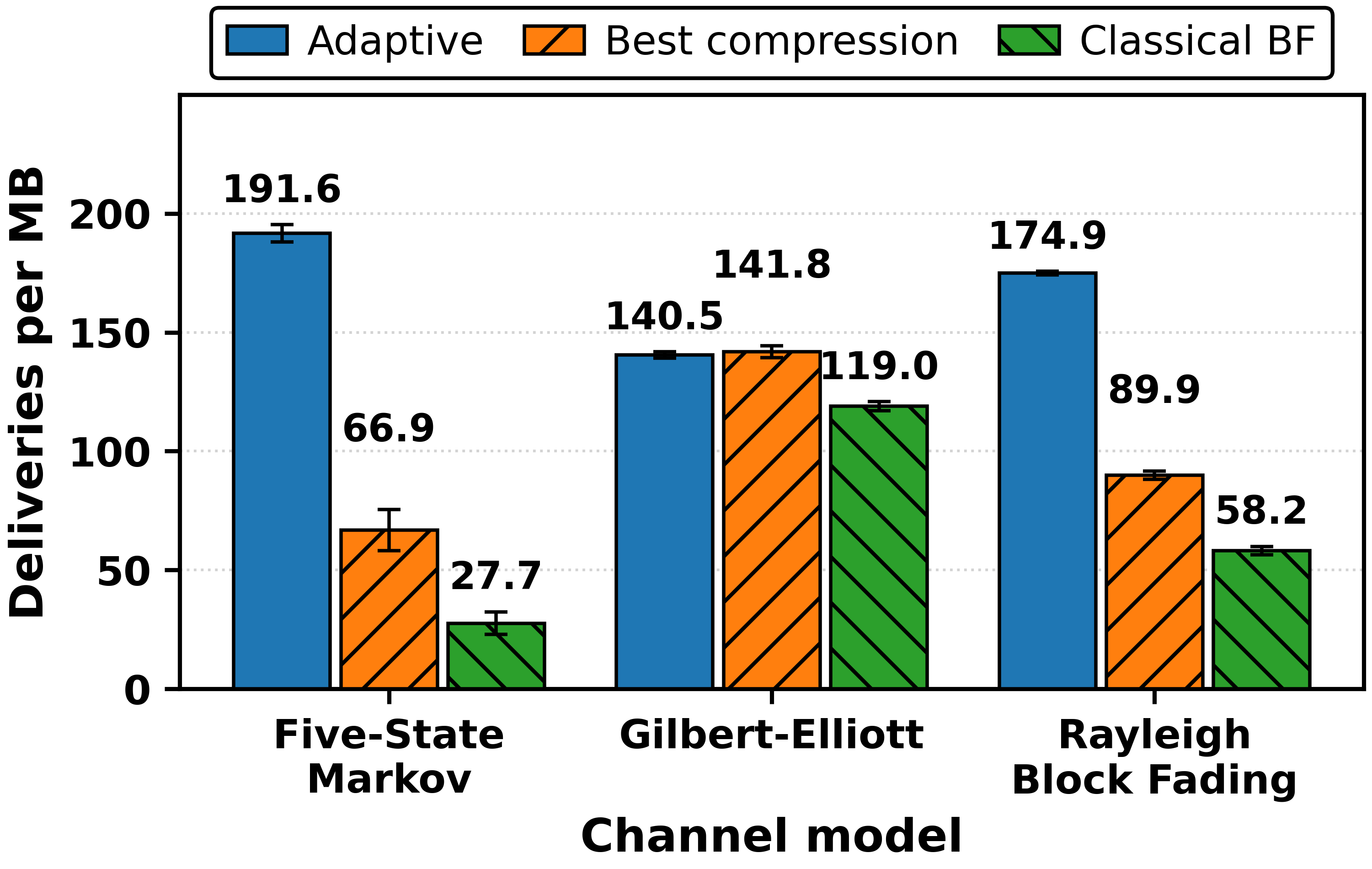}
    \caption{Communication efficiency under the three channel models. Bars show mean successful deliveries per MB and 95\% confidence intervals across 20 paired channel traces; higher values indicate better performance.}
    \label{fig:communication_efficiency}
\end{figure}

Figure~\ref{fig:communication_efficiency} compares adaptive folding with the classical full filter and the best-performing lossless-compression baseline. For each dataset and channel model, the latter is the method achieving the highest communication efficiency among gzip, zlib, bzip2, and LZMA. Reporting the strongest of these methods provides a conservative comparison with adaptive folding. Relative to the 5,991-byte sparse payload, gzip produced payloads of 5,201--5,206 bytes across the two datasets, corresponding to reductions of 13.1--13.2\%; zlib produced 5,189--5,194 bytes (13.3--13.4\%); bzip2 produced 5,913--5,923 bytes (1.1--1.3\%); and LZMA produced 5,400--5,412 bytes (9.7--9.9\%).

Under the Five-State Markov Capacity model, adaptive folding achieves 191.6 successful deliveries per MB, compared with 66.9 for the strongest compression baseline and 27.7 for the classical full filter. These values correspond to improvements of approximately \(2.9\times\) and \(6.9\times\), respectively. When available capacity decreases, fixed-size representations frequently cannot be delivered, whereas adaptive folding can select a smaller FPR-qualified representation that fits the current opportunity.

Under Gilbert--Elliott packet loss, adaptive folding and the strongest compression baseline achieve similar efficiencies of 140.5 and 141.8 deliveries per MB, respectively, while the classical filter achieves 119.0. The Good state has a high packet-success probability, allowing fixed compressed reports to succeed frequently. Consequently, reducing packet count through folding provides less benefit than under explicit capacity constraints.

Under Rayleigh block fading, adaptive folding achieves 174.9 deliveries per MB, compared with 89.9 for the strongest compression baseline and 58.2 for the classical filter. This represents improvements of approximately \(1.9\times\) and \(3.0\times\), respectively. Instantaneous fading produces a range of available capacities, allowing the adaptive policy to exploit transmission opportunities that cannot accommodate either full or fixed compressed representations.

Across the two datasets, adaptive folding achieved these efficiencies with mean selected-view FPRs of 4.36\%, 3.43\%, and 3.34\% under the Five-State, Gilbert--Elliott, and Rayleigh models, respectively; every selected view remained at or below 10\% and had zero observed false negatives.

\subsection{Receiver Freshness}
\begin{figure}[t]
    \centering
    \includegraphics[width=0.90\columnwidth]{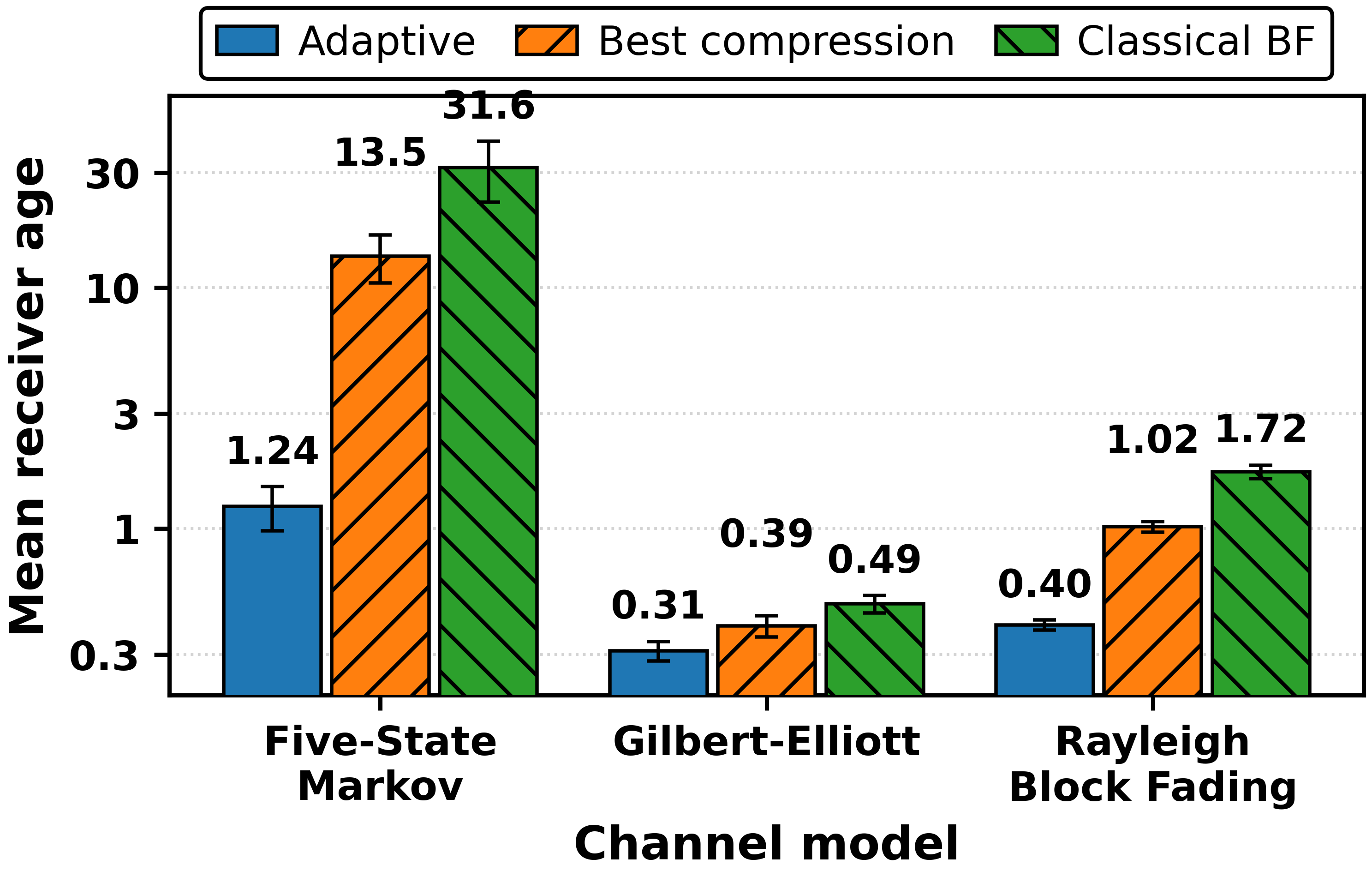}
    \caption{Mean receiver report age under the three channel models. Bars show means and 95\% confidence intervals across 20 paired channel traces; the vertical axis is logarithmic and lower values indicate fresher receiver state.}
    \label{fig:receiver_age}
\end{figure}

Figure~\ref{fig:receiver_age} reports receiver freshness as the number of transmission-opportunity intervals since the most recent successful update. Lower mean age indicates that the receiver holds a more recent representation.

Under the Five-State Markov Capacity model, adaptive folding reduces mean receiver age to 1.24 intervals, compared with 13.5 for the strongest compression baseline and 31.6 for the classical filter. This represents reductions by factors of approximately \(10.9\) and \(25.5\), respectively.

Under the Gilbert--Elliott model, the strongest compression baseline achieves marginally higher communication efficiency, but adaptive folding produces the lowest mean age: 0.31 intervals, compared with 0.39 for compression and 0.49 for the classical filter. The adaptive policy can reduce the number of packets in a report when doing so improves its expected complete-report delivery probability.

Under Rayleigh block fading, adaptive folding achieves a mean age of 0.40, compared with 1.02 for the strongest compression baseline and 1.72 for the classical filter. Receiver age is therefore reduced by approximately 61\% relative to compression and 77\% relative to the classical filter.

\subsection{Results Across Datasets and Channel Models}
The relative behavior is consistent across the two URL datasets. Adaptive folding provides its largest gains under the Five-State Markov Capacity and Rayleigh models, where the resources available to an individual reporting opportunity vary substantially. Under the more favorable Gilbert--Elliott parameters, it remains comparable in communication efficiency while producing the lowest receiver age.

These results do not show that folding universally outperforms lossless compression. Rather, they identify the conditions under which channel-aware selection is useful. Fixed lossless compression is effective when a single representation can be delivered reliably, whereas channel-aware folding is most beneficial when communication opportunities vary and smaller FPR-qualified representations can prevent missed updates. The canonical filter and inserted set remain unchanged; adaptation affects only the transmitted representation and the query-coordinate mapping associated with its retained length.

%%%%%%%%%%%%%%%%%%%%%%%%%%%%%%%%%%%%%%%%%%%%%%

\section{Discussion}
\label{sec:discussion}

\subsection{Relationship to Prior Folding}
OR folding is established prior art~\cite{broder2004network,sailhan2012folding}. Sailhan and Stehr~\cite{sailhan2012folding} select among integer-factor representations as cardinality changes while maintaining a prescribed FPR range; Chen et al.~\cite{chen2020efficient} similarly select reduced sizes from element count and maximum FPR. These approaches respond principally to the filter's internal state.

The present experiments instead fix the inserted set, canonical length, and hash count while selecting the transmitted view from current communication conditions. At the fixed cardinality of 5,000 elements, a cardinality-driven controller has no changing set-size signal; the complete-filter baseline represents its fixed operating point only if the selected FPR target requires that size. This is not a complete evaluation of the Sailhan--Stehr algorithm. The mechanisms could be combined in a two-stage controller that first selects an operative size from cardinality and FPR requirements and then selects a transmitted view from the communication opportunity.

\subsection{Folding, Compression, and Deployment}
Lossless compression produces one exactly recoverable payload for a given bit array, whereas folding exposes multiple directly queryable sizes with different FPRs. Folding is useful when a fixed compressed payload cannot be supported but a smaller FPR-qualified view can. The methods could be combined, although increased occupancy may reduce compressibility; this combination was not evaluated.

In deployment, the sender may generate or cache folded views. Candidate ratios should reflect the application's FPR tolerance and anticipated communication conditions, and only qualified representations should be retained. If a fixed compressed payload is consistently supportable, adaptive folding is unlikely to justify its selection logic and metadata.

\subsection{Limitations and Future Work}
The controlled design uses two URL datasets at the same cardinality with fixed contents. This isolates channel-driven selection but does not establish generality across other workloads, cardinalities, occupancies, or changing contents. The simulated channel models provide repeatable capacity-variation, burst-loss, and fading conditions on paired traces, but are normalized to the complete sparse-filter wire size. Measured traces or a physical testbed would strengthen external validity. The controller also assumes timely communication-state information and does not model estimation error, delayed feedback, or metadata explicitly.

Fixed folded representations are not included as baselines. Such a representation could reduce communication cost and improve delivery under constrained conditions, but would incur its higher FPR even when the channel supports greater fidelity. Adaptive folding varies this tradeoff across reporting opportunities. The results therefore establish performance relative to complete-filter and lossless-compression baselines, but do not quantify the incremental benefit over the best fixed ratio.

The single sender--receiver configuration also excludes contention, retransmission, congestion, and fairness effects. Priorities for future work are direct comparison with fixed folded baselines; integration with cardinality/FPR-driven resizing~\cite{sailhan2012folding}; and evaluation with changing contents, measured channels, multiple senders, imperfect channel estimates, energy costs, downstream false-positive costs, and combined folding and compression.

%%%%%%%%%%%%%%%%%%%%%%%%%%%%%%%%%%%%%%%%%%%%%%

\section{Conclusion}
This paper evaluated a channel-aware framework that selects the largest FPR-qualified OR-folded view supported at each reporting opportunity while retaining the canonical Bloom filter. Across two phishing URL datasets and three channel models, adaptive folding improved communication efficiency and receiver freshness over complete-filter and lossless-compression baselines under the Five-State and Rayleigh models. Under Gilbert--Elliott, it remained competitive in efficiency and produced the freshest receiver state.

Smaller views trade increased FPR for fresher updates while preserving the no-false-negative property, complementing rather than replacing lossless compression. Future work should examine fixed-fold baselines, changing contents and cardinalities, measured channels, imperfect state estimates, multiple senders, energy costs, and combined folding and compression.

%%%%%%%%%%%%%%%%%%%%%%%%%%%%%%%%%%%%%%%%%%%%%%

\IEEEtriggeratref{2}

\vspace{12pt}

\end{document}